\documentclass[twocolumn,showpacs,amsmath,amssymb]{revtex4}
\usepackage{txfonts}
\usepackage{graphicx}
\begin{document}
\title{Exact solutions of embedding the 4D Universe in a 5D Einstein manifold}
\author{Xin-He Meng$^{1,3,4}$}
\email{xhm@nankai.edu.cn}
\author{Jie Ren$^2$}
\email{jrenphysics@hotmail.com}
\author{Hong-Guang Zhang$^1$}
\affiliation{$^1$Department of physics, Nankai University, Tianjin
300071, China} \affiliation{$^2$Theoretical Physics Division, Chern
Institute of Mathematics, Nankai University, Tianjin 300071, China}
\affiliation{$^3$BK21 Division of Advanced Research and Education in
physics, Hanyang University, Seoul 133-791, Korea}
\affiliation{$^4$Department of physics, Hanyang University, Seoul
133-791, Korea}
\date{\today}
\begin{abstract}
One of the simplest way to extend 4D cosmological models is to add
another spatial dimension to make them 5D. In particular, it has
been shown that the simplest of such 5D models, i.e., one in which
the right hand side of the Einstein equation is empty, induces a 4D
nonempty Universe. Accordingly, the origin of matter in the 4D real
Universe might be mathematically attributed to the existence of one
(fictitious) extra spatial dimension. Here we consider the case of
an empty 5D Universe possessing a cosmological constant $\Lambda$
and obtain exact solutions for both positive and negative values of
$\Lambda$. It is seen that such a model can naturally reduce to a
power law $\Lambda$CDM model for the real Universe. Further, it can
be seen that, the arbitrary constants and functions appearing in
this model are endowed with definite physical meanings.
\end{abstract}
\pacs{04.50.+h,98.80.-k} \maketitle

\section{Introduction}
It is very likely that the expansion of our Universe is currently in
an accelerating phase, supported by the most direct evidence from
the measurements of type Ia supernova, and others such as the
observations of CMB by the WMAP satellite, large-scale galaxy
surveys by 2dF and SDSS \cite{Perlmutter,Spergel,sdss}. But now the
mechanisms responsible for the accelerating expansion are not very
clear. Many authors introduce a mysterious cosmic fluid called dark
energy to explain this (see Ref.~\cite{Peebles} for a review). On
the other hand, some authors suggest that maybe there does not exist
such mysterious dark energy, but instead the observed cosmic
acceleration is a signal of our first real lack of understanding of
gravitational physics \cite{Lue}. An example is the braneworld
theory with the extra dimensions compactified or noncompactified
\cite{Dvali,rs,hw}. Many cosmological models are inspired from the
string theory, which may be incorrect at the worst and speculative
at the best \cite{woit}. Despite the apparent evidence for an
accelerated expansion, there is also some ``neglected" but important
experimental evidence against the $\Lambda$ Cold Dark Matter
($\Lambda$CDM) model, which is the mainstream model in cosmology
\cite{lieu}. Nevertheless, given the presently popular notion, it is
reasonable to study the cosmological models that can be reduced to
the $\Lambda$CDM model, as in our previous works \cite{rm} and the
present one.

Space-Time-Matter (STM) theory \cite{wes1,wes2,liu95,liu01} suggests
that our Universe is a four-dimensional (4D) hyperspace embedded in
a 5D Ricci-flat manifold, \textit{i.e.}, the matter in the 4D real
Universe is induced by a 5D Ricci-flat Universe. Leon has shown that
the braneworld theory and the induced-matter theory are actually
equivalent \cite{leon}. Solutions of the 5D Einstein equation
without any matter source can always be regarded as solutions of the
corresponding 4D gravitational field equations with the induced
matter, as the Compbell theorem indicates \cite{pw}. However, in the
previous work the 5D Universe is assumed to be Ricci flat, i.e.,
without a 5D cosmological constant. In the present work, we assume
that a cosmological constant exists in the 5D Universe as a more
general case to study the features of the model, and demonstrate
that the reduction to the 4D real Universe can be realized by a
simple choice of the arbitrary functions in the metric. At the same
time, each function and constant in the metric separately possesses
explicit physical meanings.

The paper is planned as follows. In the next section we review how a
solution of 5D Ricci-flat Universe in the induced-matter theory can
be used to generate a simple 4D cosmology. In section 3 we give the
exact solutions of the 5D Universe with a cosmological constant and
study the new features. The last section devotes discussions and
conclusions for the general framework studies to this toy model.

\section{5D Ricci-flat Universe}
Firstly, we summarize some basics of STM theory
\cite{wes1,wes2,liu95,liu01}, as a preparation for our work. STM
theory starts with a solution of Einstein's equation in 5D
Ricci-flat spacetime,
\begin{equation}
R_{ab}=0,\label{eq:rab}
\end{equation}
where the subscript indices run from 0 to 4. A new class of solution
for Eq.~(\ref{eq:rab}), which extends the FRW solutions, is given in
Ref.~\cite{liu95} (and we take their conventions hereafter, i.e., we
work in natural units where $c=8\pi G=1$):
\begin{eqnarray}
ds^2 &=& B^2dt^2-A^2\left(\frac{dr^2}{1-kr^2}+r^2d\Omega
^2\right)-dy^2,\label{eq:ds}\\
A^2(t,y) &=& (\mu^2+k)y^2+2\nu y+\frac{\nu^2+K}{\mu
^2+k},\label{eq:a}\\
B(t,y) &=& \frac{1}{\mu}\frac{\partial A}{\partial t}\equiv
\frac{\dot{A}}{\mu}.
\end{eqnarray}
Here $\mu =\mu(t)$ and $\nu =\nu(t)$ are arbitrary functions, $k$ is
the 3D curvature index ($k=\pm 1,0$) and $K$ is a constant that
relates the 5D Kretschmann invariant as the form
$R_{abcd}R^{abcd}=72K^2/A^8$, which shows that $K$ determines the
curvature of the 5D manifold. This framework has been intensely
studied in the literature \cite{liu01,zha06,xu}. The spacetime
geometry is encoded in the functions $A$ and $B$, which are
dependent on the extra dimension $y$. Since the form of $Bdt$ is
invariant under an arbitrary transformation $t=t(\widetilde{t})$,
one of the two arbitrary functions $\mu(t)$ and $\nu(t)$ can be
fixed without changing the generality of the solution. Here
$\mu(t)=0$ corresponds to a singularity, and if a $\mathbb{Z}_2$
symmetry of the metric with respect to the extra dimension $y$ is
required, the another free function $\nu(t)$ must disappear.

The corresponding 4D line element is
\begin{equation}
ds_4^2=g_{\alpha \beta}dx^\alpha
dx^\beta=B^2dt^2-A^2\left(\frac{dr^2}{1-kr^2}+r^2d\Omega^2\right),
\end{equation}
which is similar to the Robertson-Walker metric and thus underlies
the standard FRW models. We can calculate the non-vanishing
components of the 4D Ricci tensor and its corresponding scalar. The
4D Einstein tensor $^{(4)}G_\beta^\alpha\equiv
\,^{(4)}R_\beta^\alpha-\delta_\beta^\alpha\,^{(4)}R/2$ with its
non-vanishing components is
\begin{eqnarray}
^{(4)}G_0^0 &=&\frac{3(\mu^2+k)}{A^2},\\
^{(4)}G_1^1 &=&
\,^{(4)}G_2^2=\,^{(4)}G_3^3=\frac{2\mu\dot{\mu}}{A\dot{A}}+\frac{\mu
^2+k}{A^2}.
\end{eqnarray}
These give the components of the induced energy-momentum tensor
since Einstein's equation $G_\beta ^\alpha =T_\beta ^\alpha$ holds.
Provided that the induced matter is described by a perfect fluid
with density $\rho$ and pressure $p$ moving with a 4-velocity
$u^\alpha \equiv dx^\alpha /ds$, plus a cosmological term whose
nature is to be determined, the energy-momentum tensor is given by
\begin{equation}
^{(4)}G_{\alpha\beta}=(\rho+p)u_\alpha u_\beta+(\rho_\Lambda
-p)g_{\alpha\beta}.
\end{equation}

The above framework provides special cases with confrontations to
physics observations. For a braneworld model, the authors of
Ref.~\cite{liko} have studied a solution that incorporates the
$\mathbb{Z}_2$ reflection symmetry condition
$g_{\alpha\beta}(x^{\gamma},y)=g_{\alpha\beta}(x^{\gamma},-y)$,
which can be achieved directly by setting $\nu=0$ in the metric
solution. Different choices and the corresponding forms of $A(t,y)$
and $B(t,y)$ may give differently concrete models, which can be
found in Refs.~\cite{wes1,liu01,zha06,xu}. At present we are most
interested in the evolution of the Universe including the so called
``dark energy" (if it really exists without the possible new
signature that our gravity is to be modified at cosmic scale).

\section{Reduction to $\Lambda$CDM model in the 5D Universe with cosmological constant}
We assume that a cosmological constant $\Lambda$ exists in the 5D
Universe as a more general case. Starting with the ansatz of
Eq.~(\ref{eq:ds}), we solve Einstein's equation
$R_{ab}-\frac{1}{2}g_{ab}R=\Lambda g_{ab}$ and obtain the solution
for $\Lambda>0$,
\begin{eqnarray}
A(t,y)^2=\frac{2(\mu^2+k)(1-\cos\lambda
y)}{\lambda^2}+\frac{2v\sin\lambda y}{\lambda}\nonumber\\
+\frac{2[\mu^2+k-\sqrt{(\mu^2+k)^2-\lambda^2(\nu^2+K)}]\cos\lambda
y}{\lambda^2},\label{eq:main}
\end{eqnarray}
where $\lambda=\sqrt{6\Lambda}/3$, $\mu=\mu(t)$ and $\nu=\nu(t)$.
And $B(t,y)=\dot{A}/\mu$ remains unchanged. For the case
$\Lambda<0$, by defining $\lambda=\sqrt{6|\Lambda|}/3$, the solution
is
\begin{eqnarray}
A(t,y)^2=\frac{2(\mu^2+k)(\cosh\lambda
y-1)}{\lambda^2}+\frac{2v\sinh\lambda y}{\lambda}\nonumber\\
+\frac{2[\sqrt{(\mu^2+k)^2+\lambda^2(\nu^2+K)}-(\mu^2+k)]\cosh\lambda
y}{\lambda^2},
\end{eqnarray}
These solutions can be reduced to Eq.~(\ref{eq:a}) by taking the
limit $\lambda\to 0$. We will show that the positive $\Lambda$
induces a positive cosmological constant in the 4D Universe under
some conditions. For simplicity we first consider the case
$k=K=y=0$; then
\begin{equation}
A^2=\frac{2(\mu^2-\sqrt{\mu^2+\lambda^2\nu^2})}{\lambda^2}.\label{eq:lcdm}
\end{equation}
Now we substitute the simple choice of $\mu(t)$ and $\nu(t)$
\begin{equation}
\mu(t)=\dot{A},\quad\nu(t)=\sqrt{CA},\label{eq:mn}
\end{equation}
where $C$ is a parameter, to Eq.~(\ref{eq:lcdm}). After some
arrangements, this solution can be naturally reduced to the form
\begin{eqnarray}
H^2 &=& CA^{-3}+\frac{\lambda^2}{4}\nonumber\\
&=& H_0^2[\Omega_mA^{-3}+\Omega_\Lambda],
\end{eqnarray}
where $H=\dot{A}/A$, $\Omega_m=C/H_0^2$, and
$\Omega_\Lambda=\Lambda/(6H_0^2)$. This is essentially the
$\Lambda$CDM model, and the cosmological constant in the 5D Universe
contributes a term as the cosmological constant in the 4D real
Universe.

In the general case, taking $y=0$ in Eq.~(\ref{eq:main}) gives
\begin{equation}
A^2=\frac{2[\mu^2+k-\sqrt{(\mu^2+k)^2-\lambda^2(\nu^2+K)}]}{\lambda^2}.\label{eq:a2}
\end{equation}
By substituting Eq.~(\ref{eq:mn}) in Eq.~(\ref{eq:a2}), we obtain
\begin{eqnarray}
H^2 &=& CA^{-3}-kA^{-2}+KA^{-4}+\frac{\lambda^2}{4}\nonumber\\
&=&
H_0^2[\Omega_mA^{-3}-\Omega_kA^{-2}+\Omega_KA^{-4}+\Omega_\Lambda],
\end{eqnarray}
where $\Omega_k=k/H_0^2$ and $\Omega_K=K/H_0^2$. It turns out that
the constant $K$ contributes a term that describes radiation.
Conventionally, the redshift is defined by $z=1/A-1$, thus
$A^{-1}=1+z$. Compared with the expression of $H^2$ for the power
law $\Lambda$CDM model, we can see that from this 5D Universe with
cosmological constant, the arbitrary functions and constants in the
metric are endowed with explicit physical meanings, which are
summarized in Table~\ref{tab:t1}.

\begin{table}[h]
\caption{\label{tab:t1} Physical meanings of functions and constants
in the metric of the 5D Universe}
\begin{ruledtabular}
\begin{tabular}{ccc}
Functions or constants & Physical meanings & Terms in $H^2$\\
\hline
$\mu(t)=\dot{A}$ & Velocity &\\
$\nu(t)=\sqrt{CA}$ & Matter (dust) & $\Omega_m(1+z)^3$\\
$k$ & Curvature & $\Omega_k(1+z)^2$\\
$K$ & Radiation & $\Omega_K(1+z)^4$\\
$\Lambda$ & Cosmological constant & $\Omega_\Lambda$
\end{tabular}
\end{ruledtabular}
\end{table}

The reduction to the $\Lambda$CDM model can be also realized in 5D
Ricci-flat Universe by the choice
\begin{equation}
\mu(t)=\dot{A},\quad\nu(t)=\tilde{H}(z)A^2,
\end{equation}
where $\tilde{H}(z)$ describes the $\Lambda$CDM model as
\begin{equation}
\tilde{H}(z)^2=H_0^2[\Omega_m(1+z)^3+1-\Omega_m].
\end{equation}
Thus, the general case with $y=0$ gives
\begin{equation}
H(z)^2=\tilde{H}(z)^2-k(1+z)^2+K(1+z)^4.
\end{equation}
Of course we can encode more physics in the function $\nu(t)$, but
in the 5D Universe with cosmological constant, each arbitrary
function and constant separately possesses physical meanings, which
is more elegant.

\section{Conclusion}
We have presented the exact solutions of embedding the 4D Universe
in a 5D Einstein manifold, and investigated the 5D cosmological
model with cosmological constant and the reduction to the power law
$\Lambda$CDM model for the 4D real Universe. As a generalization of
the 5D Ricci-flat solution in STM theory, the 5D solution with
cosmological constant has new phenomenological features. Each
function and constant in the metric can separately possess explicit
physical meanings when we perform a 4D reduction. We can also study
the general properties of the induced matter as the contents in our
Universe, without specifying the form of the arbitrary functions.
The phantom case can also be realized, for example, the equation of
state parameter $w=p/\rho<-1$, if we take $\lambda=k=\nu=0$ and
$K=1$. We think this picture is in conformity with other models to
explain the late-time accelerating expansion of our Universe, thus
it is worth of further endeavors.

\section*{Acknowledgements}
J.R. thanks Prof. Liu Zhao for helpful discussions. X.-H.M. is
supported partly by NSFC under No. 10675062 and partly by the 2nd
stage Brain Korea 21 Program.

\end{document}